\documentclass[12pt]{elsarticle}
\usepackage{amssymb,multirow,epstopdf,xcolor,hyperref}
\usepackage{graphicx}
\usepackage{bm}

\usepackage{hyperref}

\usepackage[normalem]{ulem}
\usepackage{color}
\definecolor{linkcolor}{HTML}{799B03}
\definecolor{urlcolor}{HTML}{799B03}

\hypersetup{linkcolor=linkcolor, urlcolor=urlcolor, colorlinks=true}

\newcommand{\br}{\boldsymbol{ r}}

\newcommand{\bu}{ \boldsymbol{u}}

\newcommand{\dd}{\mathrm{d}}
\newcommand\Ray{\mbox{\rm{Ra}}}  
\newcommand\Pran{\mbox{\rm{Pr}}} 
\newcommand\Rey{\mbox{\rm{Re}}} 

\journal{International Communications in Heat and Mass Transfer}

\begin{document}

\begin{frontmatter}

\title{Transient flows and reorientations of large-scale convection in a cubic cell}

\author[icmm]{A. Vasiliev}
\author[icmm]{P. Frick}
\author[kanpur]{A. Kumar}
\author[icmm]{R. Stepanov} 
\author[icmm]{A. Sukhanovskii\corref{cor1}}
\ead{san@icmm.ru}
\author[kanpur]{M. K. Verma}

\cortext[cor1]{Corresponding author}

\address[icmm]{Institute of Continuous Media Mechanics, Academ. Korolyov, 1, Perm, 614013, Russia}
\address[kanpur]{Department of Physics, Indian Institute of Technology, Kanpur, 208016, India}

\begin{abstract}
The transient processes of a turbulent large-scale convective circulation (LSC) in a cubic cell are investigated using large-eddy simulations for Rayleigh number  $\Ray=10^8$ and Prandtl number $\Pran=0.7$. For the first time, we have explicitly shown that LSC is accompanied by large-scale azimuthal flows with non-zero total angular momentum. It is also shown that solid-body rotation of the entire fluid is not realized. It is found that correlation between rotation of LSC plane and the mean azimuthal motion is high during quasiperiodic oscillations of LSC near the diagonal plane and relatively weak during LSC reorientations. We propose a new plausible scenario for the reorientations of the LSC in a cube that does not involve a mean azimuthal flow. Instead of  a single-roll, we introduce the superposition of a pair of large-scale orthogonal quasi-two-dimensional (Q2D) rolls and the reorientation of the LSC occurs as a result of the cessation of one of the Q2D rolls. This scenario is consistent with all known experimental and numerical data.
\end{abstract}


\end{frontmatter}


\section{Introduction}
\label{sec:intro}

Thermal convection in closed volumes of different shapes has many interesting features including variety of patterns and complex temporal dynamics~\citep{Ahlers:RMP2009,Lohse:ARFM2010,Chilla:EPJE2012,Bodenschatz:ARFM2000,Verma:book:BDF}. One of the most interesting phenomenon is the formation of large-scale circulation (LSC), known in turbulent convection as a mean wind~\citep{Sreenivasan:PRE2002, Brown:JFM2006, Xi:PRE2007}.

\citet{Sreenivasan:PRE2002}, \citet{Brown:JFM2006}, \citet{Xi:PRE2007}, and \citet{Mishra:JFM2011} performed experiments and numerical simulations on thermal convection at high Rayleigh number in a cylindrical geometry and observed that {the plane containing the LSC exhibits random drift in the azimuthal direction.  This phenomenon is called {\em flow reorientation}. Researchers argue that the flow reorientations in a cylinder occur due to two reasons: (a) {\em rotation-led reorientation}, which originates from an azimuthal reorientation of the LSC, and (b) {\em cessation-led reorientation}, in which the largest roll disappears temporarily during the reorientation} and two secondary rolls become significant.  The primary roll reappears in a new direction that is not correlated to the earlier roll direction. Long-time experimental realizations showed that the azimuthal motion consists of erratic fluctuations and a time-periodic oscillation. For the most of the time, the orientation of the wind has a preferred direction, with all other orientations appearing as transient states \citep{Xi:PRE2006}.

There have been only a few studies of LSC dynamics in a cube. At moderate values of Rayleigh number (up to $\Ray=1.58\times 10^{7}$) \citet{Gallet:GAFD2012}, showed that the dominant orientation of the LSC is perpendicular to one of  the sidewalls. After $\Ray$ exceeds some critical value $\Ray\simeq 4.3\times 10^{6}$, reorientations in which the LSC changes direction by $\pm 90^\circ$ were observed. It was proposed that the change of orientation is a result of competition between two perpendicular large-scale rolls.  \citet{valencia2007turbulent}  showed in an experimental and numerical study that in the range  $3\times 10^{7}<\Ray<10^{8}$, the LSC is a single roll with a predominant diagonal orientation. From their numerical simulation, they also claimed that the LSC changes orientation due to the rotation of a single roll around the vertical axis. At higher Rayleigh number $\Ray=5\times 10^{8}$, the dominant LSC orientation was observed along one of the diagonals \citep{Zimin1978}. Reversals of the LSC along one of the diagonals and its reorientation to the other diagonal were noted by \citet{liu2009heat}. A detailed study of turbulent convection in a cubic cell by \citet{Vasiliev:IJHMT2016} showed that the LSC switches from one to the other diagonal randomly.  \citet{bai2016} and \citet{Foroozani:PRE2017} observed similar dynamics of LSC in an experiment and in a numerical simulation, respectively.

The understanding of the observed LSC reorientation as a gradual rotation needs to be clarified. The rotation of an LSC can be considered as motion of the entire structure, which implies the appearance of an azimuthal flow. Another approach assumes that this rotation occurs as a change of the LSC direction, namely, as a rotation of the LSC plane only. The concept of the azimuthal motion of the LSC derives from \citet{Cioni:JFM1997}. From the temperature measurements made at the top and bottom of a cell, the authors proposed that the LSC moves azimuthally as a single structure.  Though this assumption is generally  implicitly or explicitly accepted,  there is no direct evidence of global azimuthal flows based on velocity measurements or numerical simulations, either in a cylinder or in a cube.
So far, the existence of an azimuthal flow is a key element of the suggested physical interpretations and mechanisms for random LSC reorientations in a cylinder and the diagonal switching of the LSC in a cube \cite{Brown:PRL2007, Brown:PF2008, brown2008azimuthal, bai2016, Foroozani:PRE2017}. Hence, it raises a question about the large-scale volumetric force required to produce a mechanical rotation of the LSC as a single structure. In \cite{brown2006effect,Brown:PRL2007, Brown:PF2008,brown2008azimuthal}, two candidates forces were proposed: the mean effect of stochastic turbulent fluctuations and  the Earth's Coriolis force. However, some  {aspects of dynamics of the flow during reorientation are still} open for discussion.

In this paper, we consider the LSC reorientations in Rayleigh--B\'enard convection in a cube and provide numerical data that explicitly show the appearance of the large-scale azimuthal flows and its relation to a rotation of the LSC plane. We propose a plausible scenario  {describing} reorientations of the LSC in a cube without azimuthal rotating flow.

\section{Numerical simulation}
\label{sec:NumSimul}

Numerical simulations of turbulent Rayleigh--B{\'e}nard convection in a cubic cell were carried out  using open-source software OpenFOAM 4.0. The discretization of the 3D Navier--Stokes equations for  incompressible flows in Boussinesq approximation is based on the method of control volumes. For turbulence modelling, we employ the large eddy simulation  approach,  namely  the  Smagorinsky--Lilly model with the Smagorinsky constant $0.14$ and turbulent Prandtl number $0.9$.  For the non-dimensionalization, we have used the cube edge $L$ as the  length  scale, $U_{f}=\sqrt{\beta g L \Delta_T }$ (free-fall velocity) as the velocity scale, and $\Delta_T$ (  {temperature difference between horizontal isothermal plates}) as the temperature scale.  {Governing parameters for Rayleigh--B{\'e}nard convection are the Rayleigh number $\mathrm{Ra} = (\alpha g L^3 \Delta_T )/(\nu \kappa)$ and Prandtl number
$\mathrm{Pr}  = \nu/\kappa$, where $g$ is the acceleration due to gravity; $\nu,\kappa,\alpha$ are respectively kinematic viscosity, thermal diffusion coefficient, and thermal expansion coefficient.}

We employed the no-slip boundary condition for the velocity at all walls. For the thermal field, adiabatic conditions are used for the vertical walls, and isothermal conditions for the horizontal plates. {$z$-axis is the vertical axis which is oriented againts the gravity vector.} Gaussian finite volume integration is used to compute the spatial derivatives, and a second-order Crank--Nicolson scheme is used for the time stepping.   {The value of Reynolds number $\Rey$ is estimated using as a characteristic velocity of LSC the maximal value over the mean velocity field, calculated for the time period between two consecutive  reorientations} and for this run $\Rey$ is approximately $3000$. We employed a constant time step $10^{-3}$ for which the Courant number is less than 0.4, hence our simulations are well resolved in time.

 {\begin{table}
\caption{Simulation parameters for $\Ray=10^8$ and $\Pran=0.7$. $N_{x}$, $N_{y}$ and $N_{z}$ are number of grid points along $x$, $y$, and $z$ directions; $\Delta_{min}^z$, $\Delta_{max}^z$ are the minimum and maximum grid spacing in the vertical $z$ direction; $N_{BL}$ is the number of grid points in the thermal boundary layer.}
\label{tab:1}
\begin{center}
\begin{tabular}{c|c|c|c|c|c}
\hline
\hline
Reference & $N_{x}\times N_{y}\times N_{z}$ & $\Delta_{min}^{z}/L$ & $\Delta_{max}^{z}/L$ & $N_{BL}$ & $Nu$ \\
\hline
Present work & $64\times 64\times 64$ & $1.5\times 10^{-3}$ & $2.4\times 10^{-2}$ & 4 & 35.7 \\
Present work & $128\times 128\times 128$ & $7.7\times 10^{-4}$ & $1.2\times 10^{-2}$ & 9 & 31.4 \\
LES ~\cite{Foroozani:PRE2017} & $64\times 96\times 64$ & $1.7\times 10^{-3}$ & $1.6\times 10^{-2}$ & 6 & 31.6 \\
DNS ~\cite{Kaczorowskixia2013} & $290\times 290\times 290$ & -- & -- & 7 & 31.3 \\
\hline
\hline
\end{tabular}
\end{center}
\end{table}}

The longest run was performed at $\Ray=10^8$ and $\Pran=0.7$  for $24000$ free-fall time units ($t_{f}=L/U_{f}$). The grid resolution of our simulation was $64^3$ in a nonuniform mesh with a higher grid concentration near the boundaries in order to resolve the boundary layer. We used short runs with $128^3$ mesh elements to confirm the convergence. { The characteristics of the mesh and their comparison with similar studies are provided in Table~\ref{tab:1}.} The general structure of the flow and temperature distribution are in good agreement with the results of the numerical simulations by \citet{Foroozani:PRE2017}.

\section{Transient flow and reorientations of LSC}
\label{sec:TransFlow}

Our main interest is a transition of the large-scale circulation.
{ The global angular momentum can be efficiently used as a measure of the large-scale rotation (for 2D convection, e.g. \citet{molenaar2004angular, castillo2016reversal}). So we suggest to quantify characteristics of the LSC through the 3D angular momentum}:
\begin{equation}
\label{moment}
\boldsymbol{\Omega}(t)= I^{-1}\int_V \br_0\times\bu(t) \, \dd V,
\end{equation}
where $\br_0$ is the radius vector relative to the cube center and $I = 1/8$ is the  moment of inertia about the   {rotation} axis of an inscribed cylinder with radius $1/2$. The horizontal components  $\Omega_x$ and $\Omega_y$ characterize LSC orientational angle $\psi=\arctan(\Omega_y/\Omega_x)$  while $\Omega_z$ is the angular velocity of the rigid rotation { of the fluid} around the vertical axis.
Then we can determine angular velocity of LSC reorientation  as
\begin{equation}\label{moment1}
\omega(t)\equiv \dot{\psi}(t)= \frac{\Omega_y(t) \dot{\Omega_x}(t) - \Omega_x(t) \dot{\Omega_y}(t)}{\Omega_x^2(t) + \Omega_y^2(t)}.
\end{equation}
We define the accumulated angle of mean azimuthal rotation as:
\begin{equation}\label{solidbody}
\theta_z(t)=\psi(0)+ \int_{0}^{t} \Omega_z(t') \, \dd t'.
\end{equation}

The dipolar azimuthal mode of the temperature or vertical velocity is often used to characterize the dynamics of LSC~\cite{Cioni:JFM1997,Brown:JFM2006,Xi:PRE2007,Mishra:JFM2011,bai2016,Foroozani:PRE2017}.  This mode is typically  computed using the data
measured by a set of probes installed at several horizontal cross sections.  For the comparison with \cite{bai2016,Foroozani:PRE2017} we also compute { the amplitude $\delta$ and the phase $\phi$ of the dipolar mode for the vertical velocity $u_z(t,{\bf r}_i)$ at 16 equispaced points located in the middle horizontal cross section of the cube. We found out that $\phi(t)$ coincides with $\psi(t)$, so for the studying of LSC orientation either local or integral characteristics can be used.}

\begin{figure*}
\vspace{0.35cm}(a)\vspace{-0.35cm}\\
\includegraphics[width=0.99\linewidth]{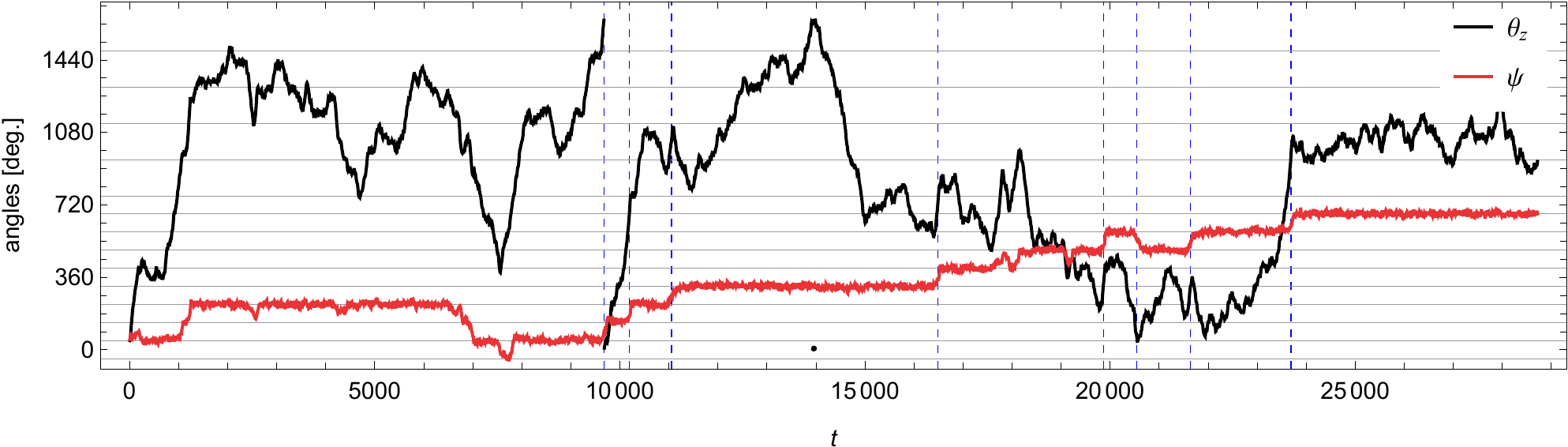}
\\\vspace{0.35cm}(b)\vspace{-0.55cm}\\
\includegraphics[width=0.99\linewidth]{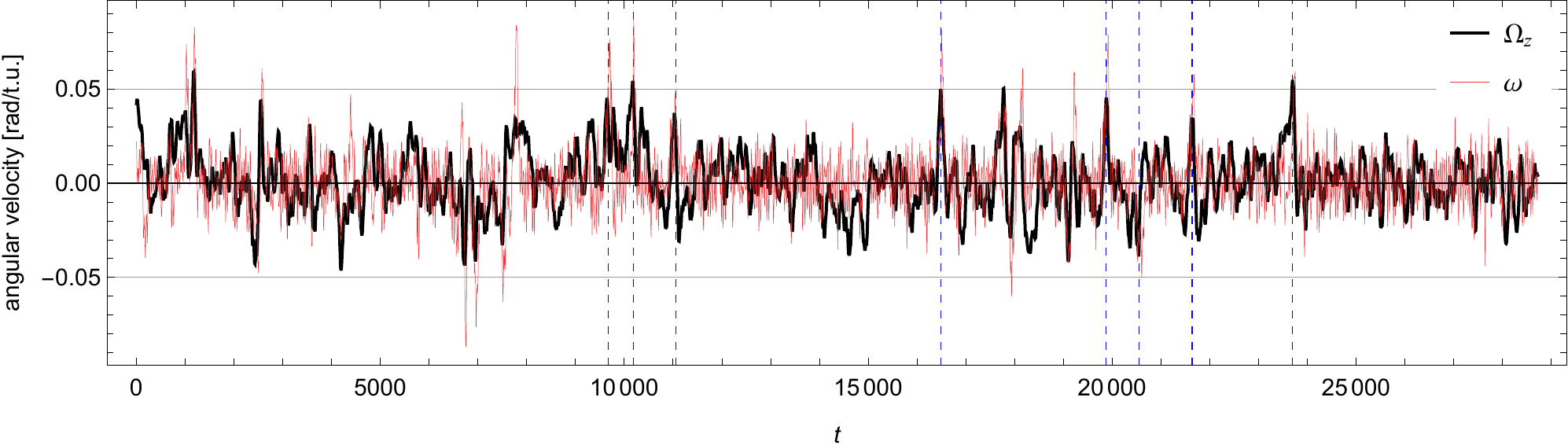}
\caption{   {(a) Time series of $\psi(t)$ and $\theta_z(t)$ ($16 \pi$ is subtracted from $\theta_z$ at $t\approx 9800$ for the better layout). (b) Time series of $\omega(t)\equiv\psi'(t)$ and $\Omega_z(t)\equiv\theta_z'(t)$ (curves are smoothed over 70 units of time).} Selected reorientation are marked by vertical dashed lines.}
\label{fig:angle_time}
\end{figure*}

Now we can clearly identify a reorientation process as {\em rotation-led reorientation} if
\begin{equation}\label{condition}
\theta_z(t)=\psi(t) \quad \textrm{and} \quad \Omega_z(t)=\omega(t).
\end{equation}
 {These conditions imply a rigid rotation of the entire fluid because the last relation becomes exact under substitution of $\Omega_x(t)= c \sin{\Omega_z t}$ and $\Omega_y(t)= c \cos{\Omega_z t}$ in Eq.~(\ref{solidbody}) (assuming that $c$ and $\Omega_z$ are constants).} The full time series of $\psi$ and $\theta_z$ are shown in Fig.~\ref{fig:angle_time}(a). The jumps of $\psi$ by $90^\circ$ correspond to the reorientations of the diagonal roll and they are like those presented in \cite{bai2016, Foroozani:PRE2017}. This proves that the specific properties of the LSC dynamics, such as reorientations, are very robust and can be reproduced using different experimental realizations and numerical simulations. Evolution of $\theta_z$ reveals rather chaotic behaviour which is intermitted by monotonic changes about several full turns. For example, such huge drop is observed near $t=14500$ but reorientation of the LSC is not happened ($\psi$ does not jump). Time derivatives (angular velocities $\omega$ and $\Omega_z$) are shown in Fig.~\ref{fig:angle_time}(b). Obviously conditions (\ref{condition}) are not realized. However we note that some of pronounced peaks  $\theta_z$  correspond to the selected  peaks  of $\psi$ (marked by dashed vertical lines in Fig.~\ref{fig:angle_time}).

\begin{figure}
\vspace{0.35cm}(a)\hspace{7.5cm}(b)\\\vspace{-0.45cm}\\
\includegraphics[width=0.5\columnwidth]{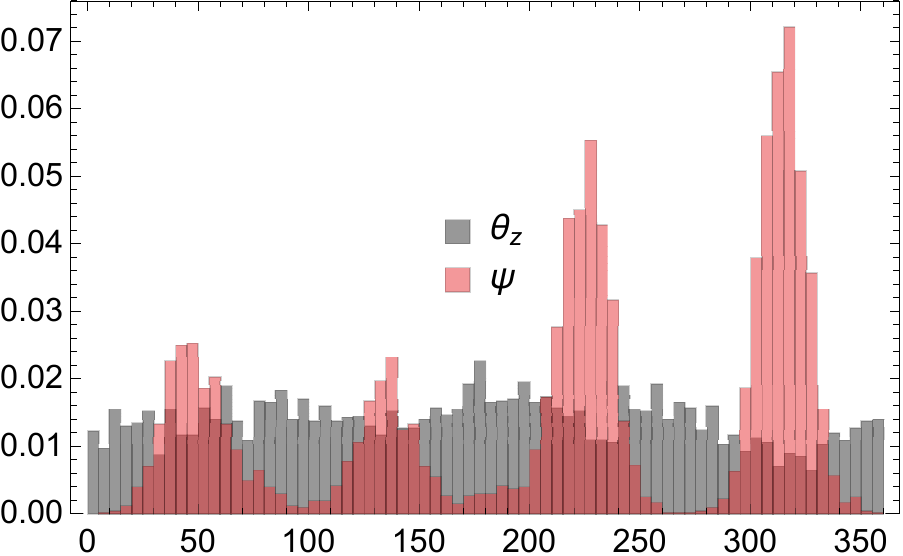}
\includegraphics[width=0.4\columnwidth]{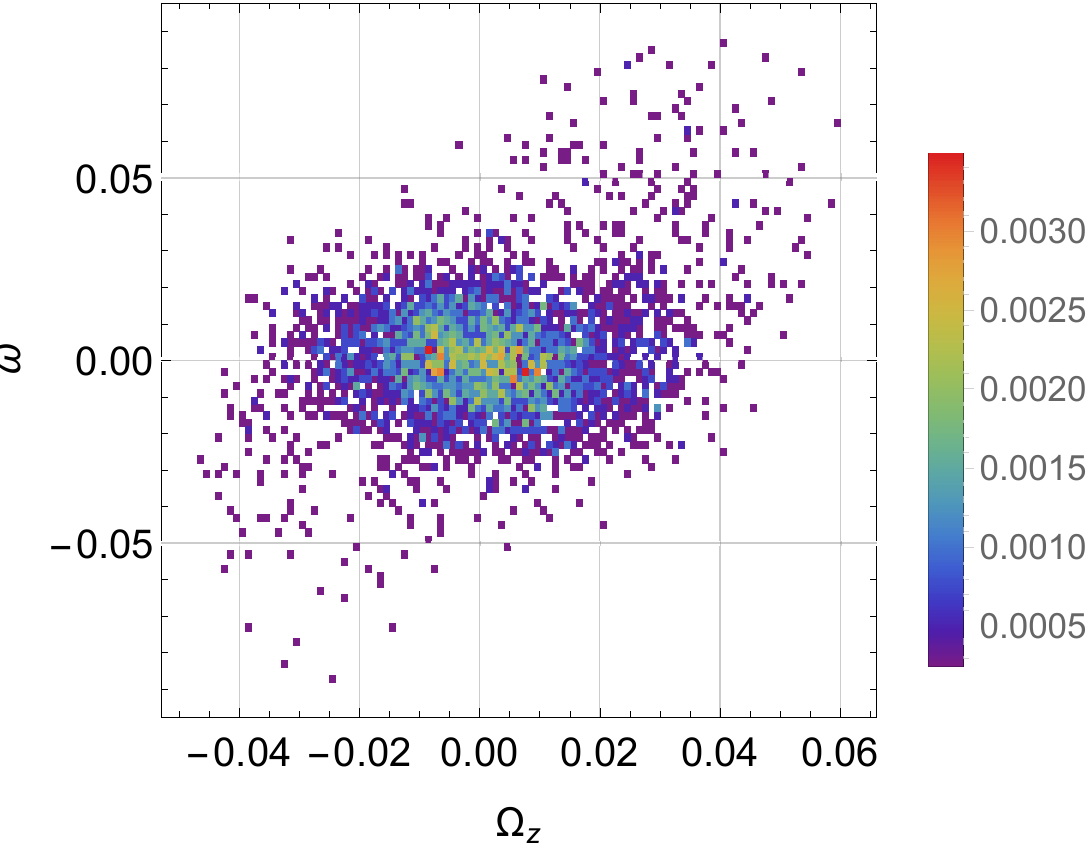}
\caption{{(a) The probability distributions of $\theta_z$ and  $\psi$  and (b) joint (two-dimensional) probability distribution of $\Omega_z$ and  $\omega$  (these signals are shown in Fig.~\ref{fig:angle_time}).}}
\label{fig:pdf}
\end{figure}

 {The valuable information can be provided by probability distributions shown in Fig.~\ref{fig:pdf}. These distributions for $\theta_z$ and  $\psi$ (Fig.~\ref{fig:pdf}(a)) clearly show that $\psi$ mostly belongs to the diagonals (angles multiple to 45 degrees) while $\theta_z$ is distributed more or less uniformly. In Fig.~\ref{fig:pdf}(b) we show joint probability of  increments of $\theta_z$ and  $\psi$ ($\omega$ and $\Omega_z$). It looks mainly symmetrical about axes which means the lack of correlations in the dynamics of two angles except relatively rare events with high magnitudes.}  Further we perform the correlation analysis to figure out the relation between the large scale azimuthal flow and the reorientation of the LSC.

Calculating the cross-correlation coefficient of two angles, $\theta_z(t)$ and $\psi(t)$, we obtained a value of $ 0.66$. However, considering the angular velocities  $\omega$ and $\Omega_z$  we have got the correlation coefficient $0.11$ only, which indicates a very weak correlation { and proves our analysis of Fig.~\ref{fig:pdf}(b)}. This result means that these two values are determined by different components of signals, namely, by short and long time variations. To quantify the role of different time-scales, we apply the wavelet analysis \cite{Mallat2008} which is an efficient tool to perform  the scale-by-scale cross-correlation analysis \cite{doi:10.1002/2015JD023265}. The continuous wavelet transform decomposes the signal $f(t)$ into map $W_f(\tau,t)$ in the time-frequency space (we use a characteristic period (time-scale) as variable) as
\begin{equation}\label{wc}
  W_f(\tau,t)=\tau^{-1} \int_{-\infty}^{\infty} f(t')\, \psi^*\left(\frac{t-t'}{\tau}\right)  dt',
\end{equation}
 {where the so-called Morlet wavelet $\psi(t)=\exp(-t^2+\imath 2 \pi t)$. }
Complex wavelet coefficients $W_f(\tau,t)$ determine an amplitude and phase of oscillation with a period $\tau$ at vicinity of time $t$.  {The global wavelet spectrum (corresponding generalization of the Fourier spectrum) is}
\begin{equation}\label{Ea}
  E_f(\tau)=\tau \int_{-\infty}^{\infty} |W_f(\tau,t)|^2 dt.
\end{equation}
The wavelet cross-correlation function $C(\tau)$ for two signals $\omega$ and $\Omega_z$ is calculated as
\begin{equation}\label{corr}
  C(\tau)=\tau (E_\omega E_{\Omega_z})^{-1/2} \int_{-\infty}^{\infty} W_\omega(\tau,t) W^*_{\Omega_z}(\tau,t) dt.
\end{equation}

\begin{figure}
\vspace{0.35cm}(a)\hspace{6.cm}(b)\\\vspace{-0.55cm}\\
\includegraphics[width=0.47\columnwidth]{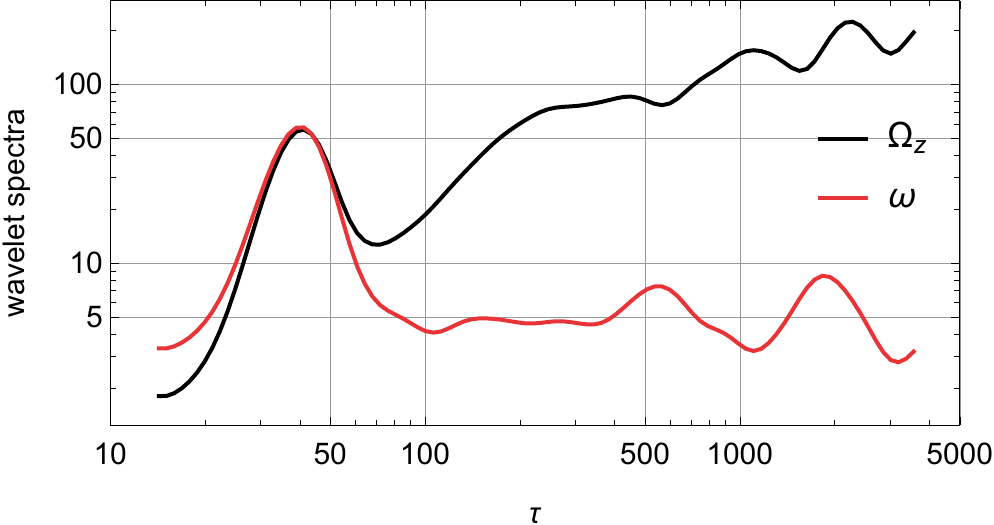}
\includegraphics[width=0.52\columnwidth]{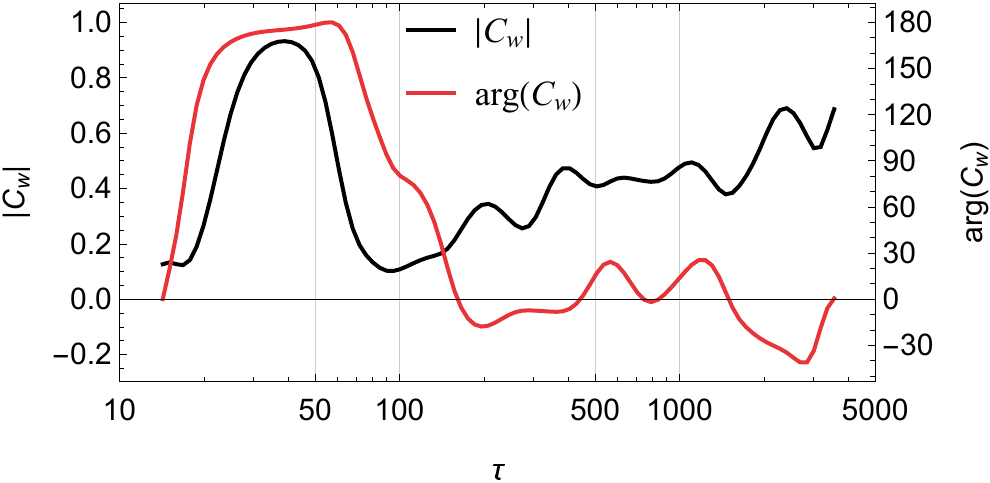}
\caption{Wavelet spectra  {(a) and the wavelet cross-correlation function (b) }for $\Omega_z(t)$ and  $\omega(t)$. { Different scales are used for modulus and phase of $C(\tau)$ in the panel (b).}}
\label{fig:wt}
\end{figure}

Wavelet cross-correlation function and individual wavelet spectra for $\Omega_z(t)$ and  $\omega(t)$ are shown in Fig.~\ref{fig:wt}. The spectra of both signals display a strong peak at $\tau \approx 40$, which corresponds to quasiperiodic oscillations of LSC near the diagonal plane.  Such oscillations with similar frequency were observed in laboratory experiments of \citet{Vasiliev:IJHMT2016}.  { $E_\omega(t)$ reaches broad plateau at $\tau \sim 200$ which represents characteristic time of reorientations (as we show it below in Fig.~\ref{fig:flips})}.  Peaks at longer time scales correspond to { inter-reorientation period and they} are seen in both spectra. The large-scale spectral power of $\Omega_z(t)$ is much stronger than the spectral power of $\omega(t)$.
The module $C(\tau)$ displays a perfect correlation at $\tau \approx 40$. To be exact, the phase of $C(\tau)$ { for this peak} corresponds to an anti-correlation. The peak of $|C|$ at $\tau\sim 200$ reveals a relatively weak correlation of $\omega$ and $\Omega_z$ during reorientation. We note that this admits $\omega\propto\Omega_z$ instead of relation (\ref{condition}).
$|C|$ slowly increases with scale at $\tau > 100$, reaching a level of about $0.7$ at time-scale of inter-reorientation period. The wavelet spectra for angles $\theta_z(t)$ and $\psi(t)$ differ from the corresponding spectra of angular velocities just by a factor of $\tau^2$, while the cross-correlation function is identical. Thus, the scale-by-scale analysis explicitly shows that correlation between $\omega$ and $\Omega_z$ is high for the time period of quasiperiodic oscillations of LSC near the diagonal plane and relatively weak for the { time period} of LSC reorientations.

To analyze the transition process in detail we consider  {velocity field in separated short time intervals near moments $t_i$ such that one of $\Omega_x$ or $\Omega_y$ changes the sign while another one does not. These eight reorientations are marked in Fig.~\ref{fig:angle_time}. Unlike the others the sixth reorientation is the clock-wise rotation. In this case we apply mirror-reflection of the velocity field relative to the horizontal plane to get $\omega(t_i)>0$ in all reorientations. Then we rotate coordinate system about the vertical axis by $90$ or $180$ degrees in order to adjust the same initial values of $\Omega_x$ and $\Omega_y$. This means that all reorientations start from the same diagonal plane.
As a result we obtain similar transition behaviour of $\boldsymbol{\Omega}$ components in all reorientations, Fig.~\ref{fig:flips} shows individual time series (thin lines) after transformations and the ensemble averages (thick lines).}
First, we analyze the horizontal components of the angular momentum. The component $\Omega_y$  decays to zero and reappears with the opposite sign. During the reorientation, the other component -- $\Omega_x$ undergoes a relatively weak enhancement. The ensemble mean of $\Omega_z$ is very weak during stable states but increases significantly (by more than its standard deviation of 0.002) during the reorientation. {Thus, we have explicitly shown statistically robust result that the reorientation is accompanied by mean azimuthal motion. However it is evident that the azimuthal motion does not necessarily lead to the reorientation of the LSC (Fig.~\ref{fig:angle_time}). It raises a question of the requirement of the large-scale azimuthal flow for reorientation of the LSC. Below we propose a plausible scenario {describing} reorientations of the LSC in a cube without azimuthal rotating flow.}

\begin{figure}
\begin{center}
\includegraphics[width=0.55\columnwidth]{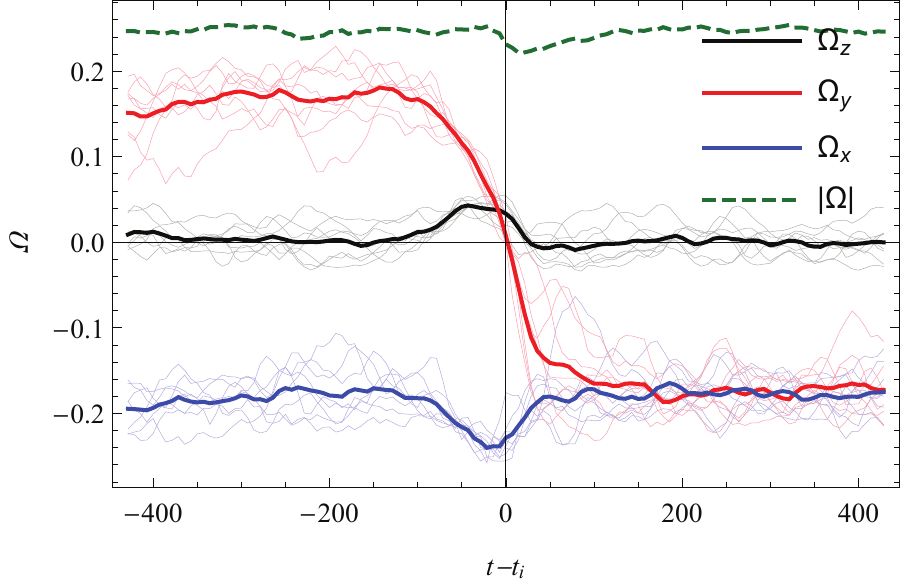}
\end{center}
\caption{Time series of $\boldsymbol{\Omega}$ components during selected reorientations (thin lines) and the ensemble averages (thick lines). Dashed line shows $|\boldsymbol{\Omega|}$.}
\label{fig:flips}
\end{figure}

\section{Non-rotational reorientation scenario}
\label{sec:NonrotatScenario}

\begin{figure}
\begin{center}
\includegraphics[width=0.6\columnwidth]{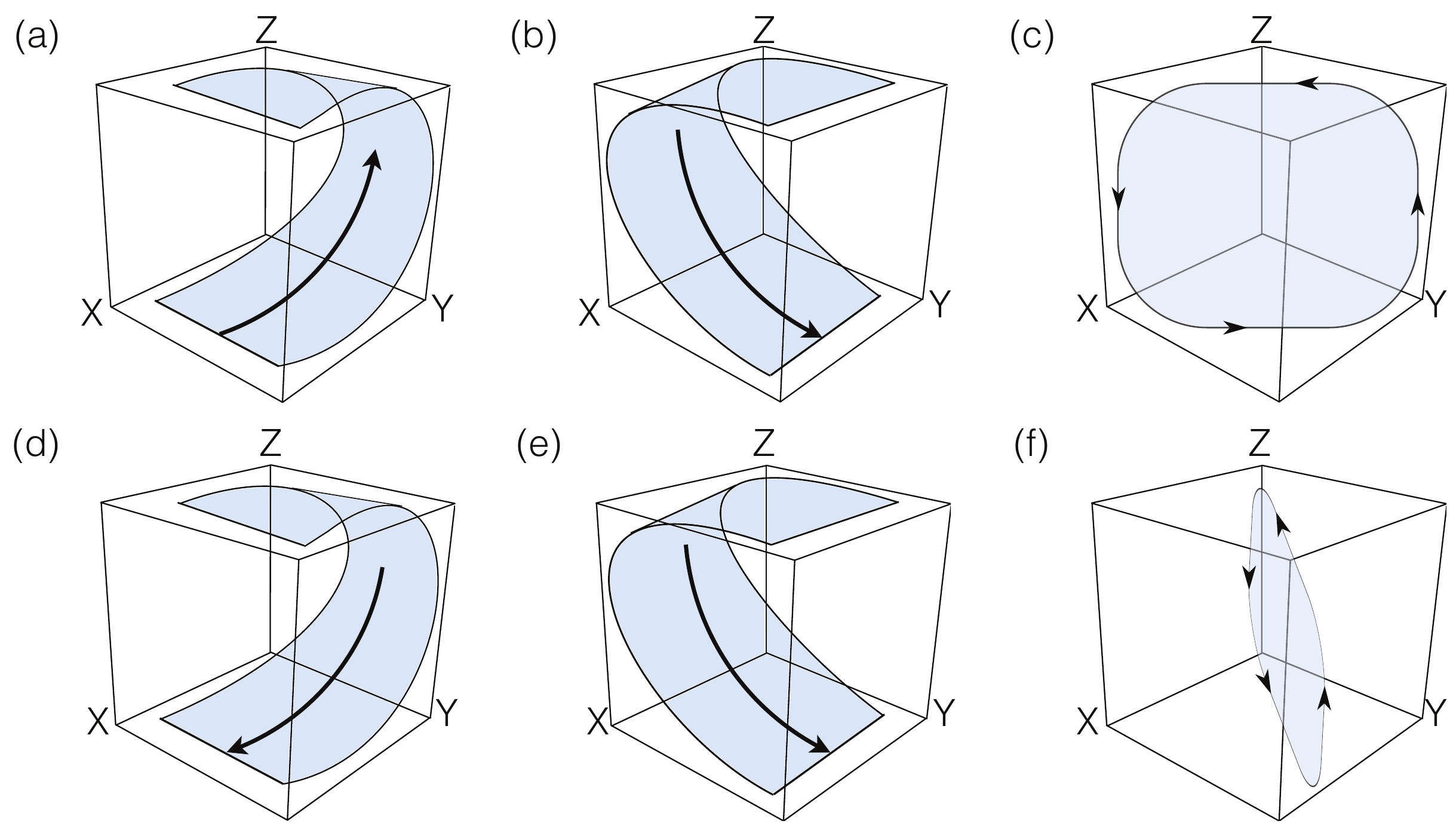}
\end{center}
\caption{Schematic diagram depicting  {that the superposition of  PLSC(a) and PLSC(b) gives the diagonal LSC(c) and the same (d) and (e) gives (f).} If one PLSC switches sign (from (a) to (d))  {and other PLSC stays the same} then LSC changes its orientation from one diagonal to the other (from (c) to (f)).}
\label{fig:schmatic}
\end{figure}

We proposed consider LSC as a result of two planar LSCs (PLSCs) in vertical orthogonal planes, $xOz$ and $yOz$ (see Fig.~\ref{fig:schmatic}).
If the PLSC in the $xOz$ plane changes direction (reverses), as shown in Fig.~\ref{fig:schmatic}(a,d), while
the PLSC in the $yOz$ plane stays in the same direction  (Fig.~\ref{fig:schmatic}(b,e)), then the diagonal LSC switches to the other diagonal (Fig.~\ref{fig:schmatic}(c,f)).
Thus, the dynamics of these PLSCs defines the orientation and intensity of the LSC.
It can be interpreted as a {\em planar cessation-led reorientation} that provides a consistent description of the LSC dynamics.

\begin{figure}
\vspace{0.35cm}(a)\hspace{7.5cm}(c)\\\vspace{-0.6cm}\\
\begin{center}
\includegraphics[width=0.49\columnwidth]{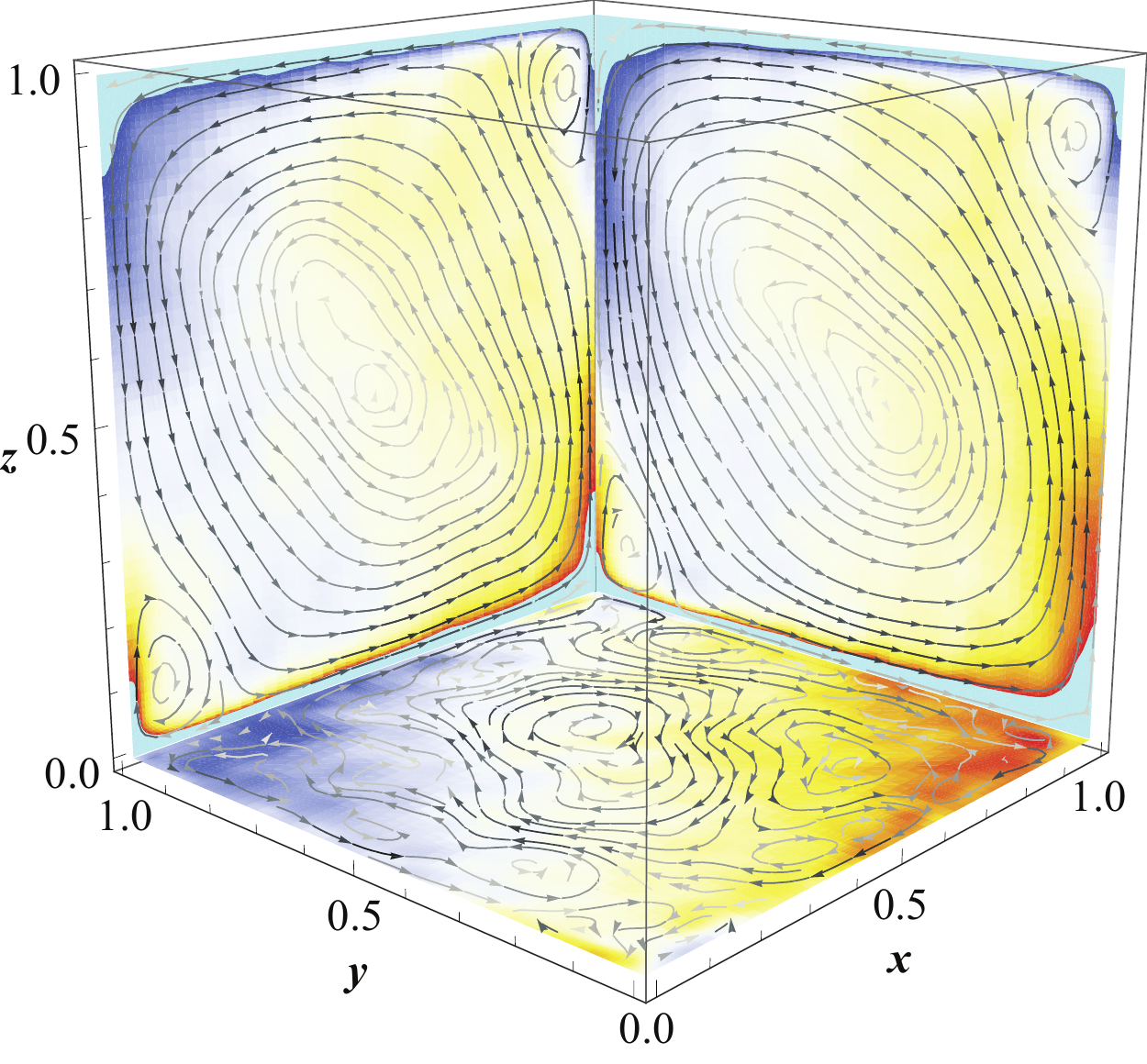}
\includegraphics[width=0.49\columnwidth]{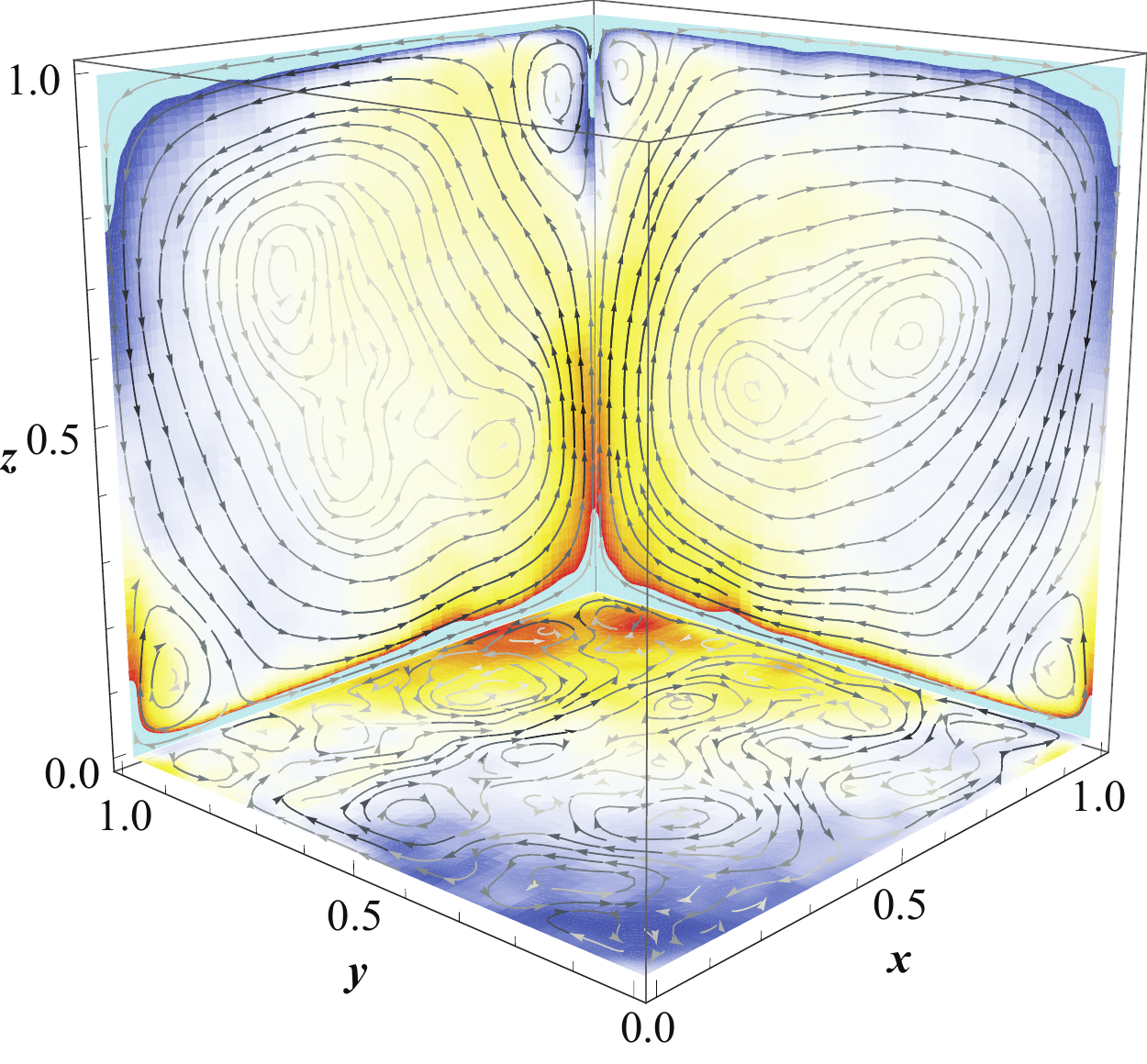}\\
(b)\includegraphics[width=0.49\columnwidth]{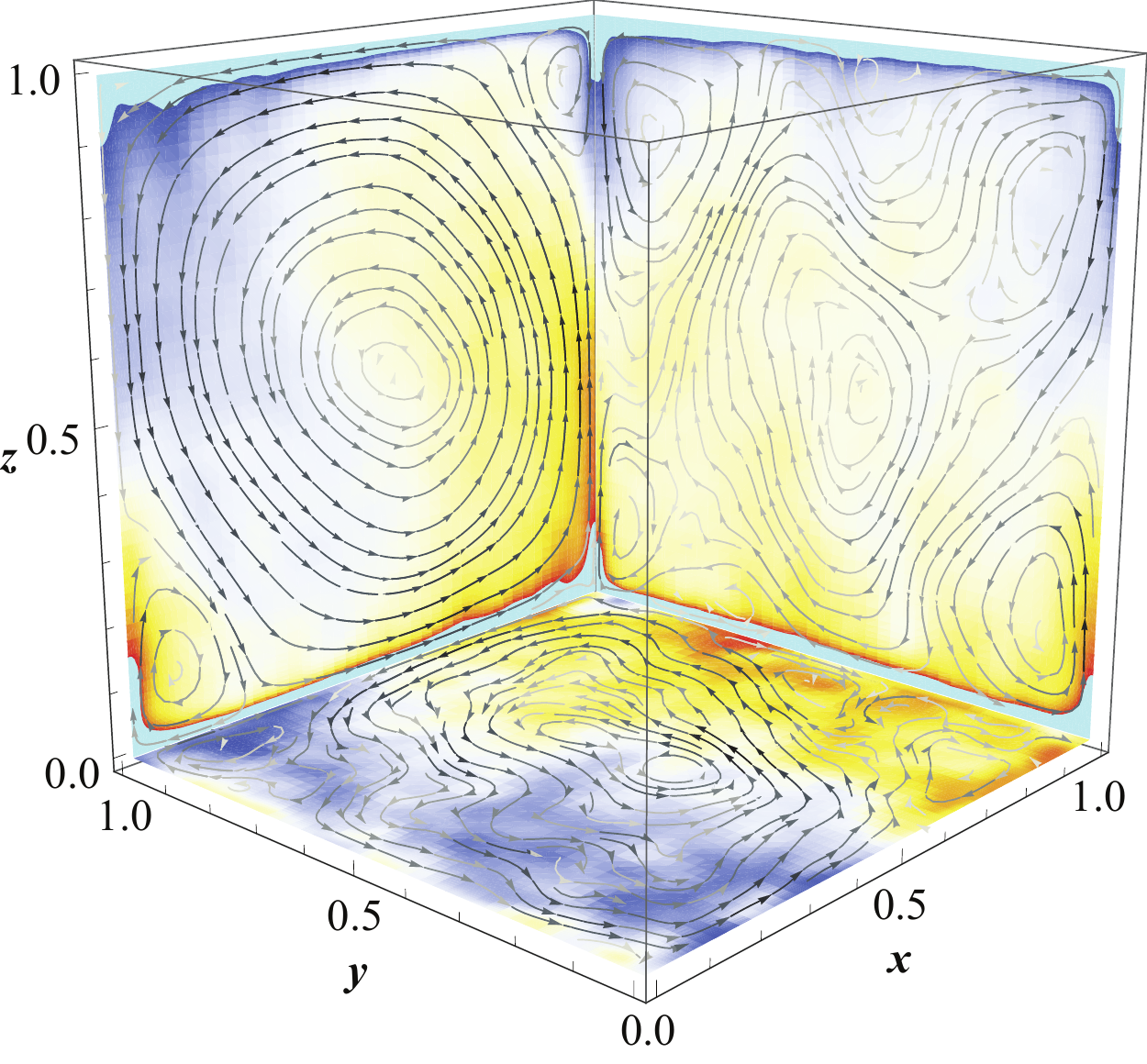}
\hspace{0.3cm}\includegraphics[width=0.11\columnwidth]{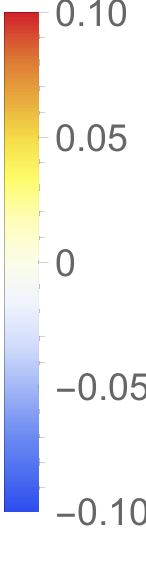}\hspace{0.15cm}
\end{center}
\caption{Velocity fields $\bar{\bu}^{\langle x\rangle}(y,z)$, $\bar{\bu}^{\langle y\rangle}(x,z)$, and $\bar{\bu}^{\langle z\rangle}(x,y)$  are shown by streamlines and temperature fields $\bar{T}^{\langle x\rangle}(y,z)$, $\bar{T}^{\langle y\rangle}(x,z)$, and $\bar{T}^{\langle z\rangle}(x,y)$ by a colour density distribution in the orthogonal planes: (a) Before reorientation, (b) during reorientation at $t=11057$,  and (c) after reorientation. Temperature  {variation} is shown in the dynamic range [$-0.1,0.1$] (the full range is [$-0.5,0.5$]). The intensity of grey of the streamlines is proportional to the velocity.}
\label{fig:cubes}
\end{figure}

To validate our proposed interpretation,
we compute the 2D velocity and temperature fields by averaging over the third coordinate.
In Fig.~\ref{fig:cubes} we present the 2D velocity fields:
\begin{eqnarray}
\bar{\bu}^{\langle x\rangle}(y,z)&=\int_0^1 \{u_y(x,y,z),u_z(x,y,z)\} \dd x, \nonumber\\
\bar{\bu}^{\langle y\rangle}(x,z)&=\int_0^1 \{u_x(x,y,z),u_z(x,y,z)\} \dd y, \\
\bar{\bu}^{\langle z\rangle}(x,y)&=\int_0^1 \{u_x(x,y,z),u_y(x,y,z)\} \dd z \nonumber
\label{threetotwo}
\end{eqnarray}
and corresponding 2D temperature fields
$\bar{T}^{\langle x\rangle}(y,z)$, $\bar{T}^{\langle y\rangle}(x,z)$ and $\bar{T}^{\langle z\rangle}(x,y)$,
to illustrate the quasi-two-dimensional (Q2D) structures before, during, and after the reorientation. {Note that angular momentum of each Q2D roll is equal to the corresponding component of $\boldsymbol{\Omega}$.}

A careful observation of Fig.~\ref{fig:cubes}  indicates that the velocity field $\bar{\bu}^{\langle y\rangle}(x,z)$ does not reverse,
but the field $\bar{\bu}^{\langle x\rangle}(y,z)$ does. So, one PLSC  vanishes and then reappears circulating in the opposite direction while the other PLSC remains stable during the reorientation. This is like the situation in the Fig.~\ref{fig:schmatic}, where the roll in the $xOz$ plane reverses, but that in the $yOz$ plane does not.{ This transformation of velocity fields holds for all observed reorientations and} it fits the {\em planar cessation-led reorientation} scenario perfectly.

Also  note  that  the reversal of the PLSC follows a very similar path as in 2D convection \cite{Chandra:PRE2011,Chandra:PRL2013, castillo2016reversal}.
The stable PLSC consists of one central roll with the largest possible scale  and two smaller counter-rotating rolls in opposite corners.
This mean-flow configuration is typical for 2D turbulent convection in square cells \cite{Sugiyama:PRL2010,Teimurazov:JCCM2012}.
During  the  cessation, the two corner rolls grow and finally reconnect to form a large-scale roll with opposite rotation. The two small-scale vortices appear at the opposite corners.

\section{Conclusions}
\label{sec:conc}

In summary, considering the transient processes of large-scale flow in a turbulent convective cubic cell, we analysed the integral characteristic  (total angular momentum) to understand the LSC dynamics. For the first time, we have explicitly shown that LSC is accompanied by spontaneous large-scale azimuthal flows with non-zero total angular momentum. Appearance of mean azimuthal motion in non-rotating turbulent Rayleigh - B\'enard convection can be very interesting for different geophysical and technological applications.
We have found that rotation of LSC plane and solid-body rotation of the entire fluid are only partly correlated at the characteristic time of reorientation process. The most of azimuthal flow spurts do not cause a reorientation of LSC. High correlation (with some phase shift) between components of total angular momentum is found for fast quasiperiodic oscillations of LSC near the diagonal. We propose a new plausible scenario for the reorientations of the LSC in a cube that does not involve a mean azimuthal flow. Instead of a single-roll, we consider the superposition of a pair of large-scale orthogonal Q2D rolls and the reorientation of the LSC occurs as a result of the cessation of one of the Q2D rolls. We emphasise similarity of the Q2D rolls transition to the known scenario of LSC reversals in the 2D convection.

\section{Acknowledgments}

This work was supported by the  Department of Science and Technology, India (INT/RUS/RSF/P-03) and Russian Science Foundation (RSF-16-41-02012) in the frame of the Indo-Russian project. The simulations have been done on the Triton supercomputer of the ICMM UB RAS, Perm, Russia. We thank for useful discussions Stephan Fauve, Joe Niemela, E. Brown and K. R. Sreenivasan.

\bibliography{Reorientations}

\end{document}